\documentclass[journal, twoside]{IEEEtran}

\usepackage{amsmath}
\usepackage{dsfont}
\usepackage{amssymb,latexsym}
\usepackage{amsfonts}
\usepackage{amscd}
\usepackage{algorithm}
\usepackage{algorithmic}
\usepackage{multirow}
\usepackage{array}
\usepackage{changepage}
\usepackage{times}
\usepackage{float}
\usepackage{graphics,subfigure}
\usepackage{graphbox}
\usepackage[latin1]{inputenc}
\usepackage{graphicx}
\usepackage{color}
\usepackage{relsize,balance}
\usepackage{cite}
\usepackage{stfloats}
\usepackage{epstopdf}
\usepackage{setspace}

\newcommand{\bs}{\mathbf}

\newcommand{\mI}{{\bs I}}
\newcommand{\mK}{{\bs K}}
\newcommand{\mP}{{\bs P}}
\newcommand{\mQ}{{\bs Q}}
\newcommand{\mR}{{\bs R}}
\newcommand{\mX}{{\bs X}}

\newcommand{\mDelta}{{\bs \Delta}}
\newcommand{\mPhi}{{\bs \Phi}}
\newcommand{\mSigma}{{\bs \Sigma}}

\newcommand{\0}{{\bs 0}}
\newcommand{\1}{{\textbf{1}}}

\newcommand{\vk}{{\bs k}}

\newcommand{\vw}{{\bs w}}
\newcommand{\vx}{{\bs x}}

\newcommand{\hatvw}{{\widetilde{\bs w}}}
\newcommand{\vvarphi}{{\boldsymbol \varphi}}
\newcommand{\vmu}{{\boldsymbol \mu}}

\def\R{\ensuremath{\mathbb{R}}}
\def\N{\ensuremath{\mathcal{N}}}
\def\E{\ensuremath{\mathbb{E}}}

\def\MSD{\text{MSD}}
\def\sgn{\text{sgn}}

\def\tr{\text{tr}}

\begin{document}

%
\title{Transient Performance Analysis of the $\ell_1$-RLS}
%
%
%

\author{Wei~Gao,~\IEEEmembership{Member,~IEEE},
        Jie~Chen,~\IEEEmembership{Senior Member,~IEEE},
        C\'edric~Richard~\IEEEmembership{Senior Member,~IEEE},
        Wentao~Shi,~\IEEEmembership{Member,~IEEE}, and
         Qunfei~Zhang,~\IEEEmembership{Member,~IEEE}\\

\thanks{Manuscript received MM DD, 2021; revised MM DD, 2021.}
\thanks{This work was supported in part by the National NSFC under Grants 62171205 and 62171380.}
\thanks{W. Gao is with the School of Computer Science and Telecommunication Engineering, Jiangsu University, Zhenjiang 212013, China (email: wei\_gao@ujs.edu.cn).}
\thanks{J. Chen, W. Shi, and Q. Zhang are with the School of Marine Science and Technology, Northwestern Polytechnical University, Xi'an 710072, China (email: dr.jie.chen@ieee.org; swt@nwpu.edu.cn; zhangqf@nwpu.edu.cn).} 
\thanks{C. Richard is with the Universit\'e C\^ote d'Azur, CNRS, OCA, 06108 Nice, France (e-mail: cedric.richard@unice.fr).}}

%
%

\markboth{IEEE SIGNAL PROCESSING LETTERS,~VOL.~xx, NO.~x, Month~2021}%
{GAO \MakeLowercase{\textit{et al.}}: Transient Performance Analysis of $\ell_1$-RLS Algorithm}
%

\maketitle

\begin{abstract}
The recursive least-squares algorithm with $\ell_1$-norm regularization ($\ell_1$-RLS) exhibits excellent performance in terms of convergence rate and steady-state error in identification of sparse systems. Nevertheless few works have studied its stochastic behavior, in particular its transient performance. In this letter, we derive analytical models of the transient behavior of the $\ell_1$-RLS in the mean and mean-square error sense. Simulation results illustrate the accuracy of these models.
\end{abstract}


%

\section{Introduction}
\label{sec:Intro}

\IEEEPARstart{S}{parsity} aware adaptive filters have been successfully applied to a wide range of applications, e.g., echo cancellation~\cite{Benesty2019}, channel estimation~\cite{Pelekanakis20104}, and system identification~\cite{Zakharov2013}. As the recursive least-squares (RLS) algorithm achieves better performance than the least-mean-square (LMS) algorithm for time-invariant system identification~\cite{Haykin1991, Sayed2003, Diniz2013}, sparse RLS-type algorithms have attracted considerable attention.

The recursive $\ell_1$-regularized least-squares (SPARLS) algorithm was introduced using the expectation-maximization scheme~\cite{Babadi2010}. The greedy sparse RLS algorithm with exponential window exploiting the orthogonal matching pursuit was devised in~\cite{Dumitrescu2012}. The $\ell_1$-norm regularized RLS ($\ell_1$-RLS) algorithm was proposed based on the modified least-squares cost function with sparsity promoting regularization~\cite{Eksioglu2010}. By considering two possible weighted $\ell_1$-norm sparsity constraints with the cost function, two weighted $\ell_1$-RLS algorithms were presented in~\cite{Eksioglu2011IET}. They can be viewed as particular cases of the convex regularized RLS (CR-RLS) algorithm, that was derived by considering any convex function in the regularizer~\cite{Eksioglu2011}. In~\cite{Hong2017}, the cost function of the original RLS was modified by adding an adaptively weighted $\ell_2$-norm penalty resulting in the proposal of two zero-attracting RLS (ZA-RLS) algorithms. A distributed sparse RLS algorithm was also proposed for decentralized scenarios over networks in~\cite{Li2014, Cao2017, Cooperative2018Djuric}.

The zero-attracting LMS (ZA-LMS) was extensively studied in~\cite{Shi2010, Zhang2014, Chen2016SPL}, with a focus on its transient behavior. In~\cite{Das2018}, the authors analyzed the mean and mean-square deviations at steady-state of the $\ell_0$-norm regularized RLS ($\ell_0$-RLS) algorithm. In contrast, despite the superiority of the $\ell_1$-RLS algorithm, no theoretical analysis of its transient behavior has been reported so far. This may be due to the fact that there are few analyses of the transient behavior of the RLS algorithm in the literature. In addition, the analysis of the update equation in the $\ell_1$-RLS is critical from a statistical perspective. To address this concern, we start this letter by reformulating the update equation of the $\ell_1$-RLS algorithm into a concise form. This makes the analysis of its transient behavior tractable both in the mean and mean-square error sense. Simulation results validate the theoretical findings.

\textit{Notation:} $[\vx]_i$ and $[\mX]_{ij}$ denote the $i$-th entry of column vector $\vx$ and the $(i,j)$-th entry of matrix $\mX$, respectively. The superscript $(\cdot)^\top$ represents the transpose of vector or matrix. The matrix trace is denoted by $\tr\{\cdot\}$. The operator $\sgn\{\cdot\}$ takes the sign of the entries of its argument. All-zero vector of length $N$ is denoted by $\0_N$, and all-one vector of length $N$ is denoted by $\1_N$. The Gaussian distribution with mean $\mu$ and variance $\sigma^2$ is denoted by $\N(\mu, \sigma^2)$. The multivariate Gaussian distribution with mean $\vmu$ and covariance matrix $\mSigma$ is denoted by $\N(\vmu, \mSigma)$. The cumulative distribution function (CDF) of the standard Gaussian distribution is denoted by $\phi(x)$. The CDF of the multivariate Gaussian distribution is denoted by $\Phi(\vx, \vmu, \mSigma)$.

\section{The Problem and $\ell_1$-RLS Algorithm}
\label{sec:Preliminaries}

Assume that the input-output sequences are generated by an unknown time-invariant system with sparse impulse response:
\begin{equation}
    \label{eq:Model}
    y_n = \vx_n^\top \vw^\star + z_n
\end{equation}
where $\vx_n \in\R^L$ is the regression vector at time instant $n$ with positive definite correlation matrix $\mR_x=\E\{\vx_n \vx_n^\top\}$, and $\vw^\star \in\R^L$ is the sparse optimal weight vector to be estimated. The modeling error $z_n$ is assumed to be stationary, white and Gaussian with zero-mean and variance $\sigma_z^2$, and statistically independent of any other signal. Consider the batch least-absolute shrinkage and selection operator (LASSO) problem usually considered for sparse system identification~\cite{Tibshirani1996}:
\begin{equation}
    \label{eq:CostFun}
    \min_{\vw \in \R^L} \left \{ \frac{1}{2} \sum_{i=0}^{n} \lambda^{n - i} |y_i - \vw^\top \vx_i|^2 + \delta \|\vw\|_1 \right \}
\end{equation}
with $0\ll\lambda<1$ a forgetting factor, and $\delta>0$ a regularization parameter that controls the trade-off between the estimation error and the sparsity of the weight vector.

Based on the modified deterministic normal equation resulting from the subgradient vector of~\eqref{eq:CostFun}, the $\ell_1$-RLS algorithm proposed in~\cite{Eksioglu2010, Eksioglu2011IET, Eksioglu2011} is given by:
\begin{align}
    \label{eq:mkn}
    \vk_n = \frac{\mP_{n-1} \vx_n}{\lambda + \vx_n^\top \mP_{n-1} \vx_n},
\end{align}
\begin{align}
    \label{eq:L1-RLS1}
    \vw_n &= \vw_{n-1} + e_n \vk_n + \delta \Big( \frac{\lambda - 1}{\lambda} \Big) \left( \mI_L - \vk_n \vx_n^\top \right) \\
    &\hspace{+2.5cm} \times \mP_{n-1} \sgn\{\vw_{n-1}\}, \nonumber \\
    \label{eq:mP1}
    \mP_n &= \lambda^{-1}(\mP_{n-1} - \vk_n \vx_n^\top \mP_{n-1}),
\end{align}
with the instantaneous estimation error $e_n=y_n -\vx_n^\top\vw_{n-1}$. Here, matrix $\mP_n$ is defined as the inverse of the time-averaged correlation matrix $\mPhi_n$ for the input vector~\cite{Haykin1991, Sayed2003}, given by:
\begin{equation}
    \label{eq:mPhi1}
    \mPhi_n = \sum_{i=0}^n \lambda^{n-i} \vx_i \vx_i^\top + \delta \lambda^{n+1} \mI_L = \lambda \mPhi_{n-1} + \vx_n \vx_n^\top.
\end{equation}
Recursion~\eqref{eq:L1-RLS1} is too complex to be used as it is in a convergence analysis. We reformulate it in an equivalent way by inserting $\vk_n=\mP_n \vx_n$ and $\mI_L-\vk_n\vx_n^\top=\lambda \mP_n \mP_{n-1}^{-1}$ from~\eqref{eq:mkn} and~\eqref{eq:mP1}, respectively, into~\eqref{eq:L1-RLS1}. This yields:
\begin{equation}
    \label{eq:L1-RLS2}
    \vw_n = \vw_{n-1} + e_n \mP_n \vx_n + \gamma \mP_n \sgn\{\vw_{n-1}\}
\end{equation}
with $\gamma=\delta(\lambda - 1)$. This equivalent formulation of the $\ell_1$-RLS algorithm makes the following analysis easier to handle. It can be seen from~\eqref{eq:L1-RLS2} that the only difference between the RLS and the $\ell_1$-RLS lies in the rightmost zero attractor term.

\section{Transient Performance Analysis of $\ell_1$-RLS}
\label{sec:Transient}

We shall now study the transient behavior of the $\ell_1$-RLS algorithm in the mean and mean-square error sense. We define the weight error vector $\hatvw_n$ as the difference between the weight vector $\vw_n$ and $\vw^\star$, i.e.,
\begin{equation}
    \label{eq:hatvw1}
    \hatvw_n = \vw_n - \vw^\star.
\end{equation}
In essence, the analysis of the $\ell_1$-RLS consists of studying the evolution over time of the expectation of $\hatvw_n$ and its correlation matrix $\mK_n=\E\{\hatvw_n\hatvw_n^\top\}$.

Before proceeding, we introduce the following statistical assumptions for mathematical tractability.

A1: The weight error vector $\hatvw_{n-1}$ is statistically independent of the regression vector $\vx_n$.

A2: Any pair of entries $[\hatvw_n]_i$ and $[\hatvw_n]_j$ with $i\neq j$ is jointly Gaussian.

The so-called independence assumption (IA) A1 is widely used in the convergence analysis of adaptive filters~\cite{Haykin1991, Sayed2003}. Assumption A2 has been used successfully in the analysis of the ZA-LMS algorithm~\cite{Chen2016SPL} as it makes the analysis of the nonlinear sign term in \eqref{eq:L1-RLS2} tractable. We shall check the validity of A2with Henze-Zirkler's multivariate normality test~\cite{Henze1990, Mecklin2005}.

\subsection{Mean Weight Error Behavior Model}
\label{subsec:MeanAnalysis}

We focus on the mean weight error analysis of the $\ell_1$-RLS. Taking the expectation of both sides of~\eqref{eq:mPhi1} yields:
\begin{equation}
    \label{eq:mPhi2}
    \E\{\mPhi_n\} = \lambda \E\{\mPhi_{n-1}\} + \mR_x
\end{equation}
with $\mPhi_{-1}=\varepsilon^{-1}\mI_L$ and $\varepsilon$ a positive initialization parameter. This recursion will be used in the following analysis. From~\eqref{eq:Model} and~\eqref{eq:hatvw1}, the instantaneous estimation error $e_n$ can be rewritten as follows:
\begin{equation}
    \label{eq:en1}
    e_n = z_n - \vx_n^\top \hatvw_{n-1}.
\end{equation}
Subtracting $\vw^\star$ from both sides of~\eqref{eq:L1-RLS2}, using~\eqref{eq:hatvw1} and~\eqref{eq:en1}, we have:
\begin{equation}
    \label{eq:vw1}
    \begin{split}
    \hatvw_n = \hatvw_{n-1} &- \mP_n \vx_n \vx_n^\top \hatvw_{n-1} \\ &+ z_n \mP_n \vx_n + \gamma \mP_n \sgn\{ \vw^\star + \hatvw_{n-1} \}.
    \end{split}
\end{equation}
Pre-multiplying both sides of~\eqref{eq:vw1} by $\mP_n^{-1}$, then applying~\eqref{eq:mPhi1} and definition $\mPhi_n=\mP_n^{-1}$, leads to
\begin{equation}
    \label{eq:vw2}
    \mPhi_n \hatvw_n = \lambda \mPhi_{n-1} \hatvw_{n-1} + z_n \vx_n + \gamma \sgn\{ \vw^\star + \hatvw_{n-1} \}.
\end{equation}
Taking the expectation of both sides of~\eqref{eq:vw2}, considering that noise $z_n$ is statistically independent of any other signal and it is zero mean, we have:
\begin{equation}
    \label{eq:vw3}
    \E\{\mPhi_n \hatvw_n\} = \lambda \E\{\mPhi_{n-1} \hatvw_{n-1}\} + \gamma \E\big\{ \sgn\{ \vw^\star + \hatvw_{n-1} \} \big\}.
\end{equation}
Using the following approximation presented in detail in~\cite{Eweda2020}:
\begin{equation}
    \label{eq:Approx1}
    \E\{\mPhi_n \hatvw_n\} \approx \E\{\mPhi_n\} \E\{\hatvw_n\}
\end{equation}
equation~\eqref{eq:vw3} becomes:
\begin{equation}
    \label{eq:vw4}
    \begin{split}
        \E\{\mPhi_n\} \E\{\hatvw_n\} &= \lambda \E\{\mPhi_{n-1}\} \E\{\hatvw_{n-1}\} \\
        &\qquad+ \gamma \E\big\{ \sgn\{ \vw^\star + \hatvw_{n-1} \}\big\}.
    \end{split}
\end{equation}
Let $u_i\sim\N(\mu_i, \sigma_i^2)$ a Gaussian random variable. Lemma~1 presented in~\cite{Chen2016SPL} shows that:
\begin{equation}
    \label{eq:Lemma1}
    \E\big\{ \sgn\{u_i\} \big\} = 1 - 2 \phi(-\mu_i/\sigma_i).
\end{equation}
In order to evaluate the last term on the r.h.s. of~\eqref{eq:vw4}, we set the $i$-th entry of $\E\big\{ \sgn\{ \vw^\star + \hatvw_{n-1} \}\big\}$ as follows:
\begin{equation}
    \label{eq:u_i}
    [\vw^\star + \hatvw_{n-1}]_i \to u_i
\end{equation}
with
\begin{align}
    \label{eq:mu}
    [\vw^\star]_i + \E\big\{ [\hatvw_{n-1}]_i \big\} &\to \mu_i, \\
    \label{eq:sigma}
    \E\big\{ [\hatvw_{n-1}]_i^2 \big\} - \E\big\{ [\hatvw_{n-1}]_i \big\}^2& \to \sigma_i^2,
\end{align}
where $\E\big\{ [\hatvw_{n-1}]_i^2 \big\}$ can be extracted from the main diagonal entries of matrix $\mK_{n-1}$ that will be determined in the next subsection. With the above definitions, it then follows that:
\begin{equation}
    \label{eq:vw5}
    \E\{\mPhi_n\} \E\{\hatvw_n\} = \lambda \E\{\mPhi_{n-1}\} \E\{\hatvw_{n-1}\} + \gamma (\1_L - 2\vvarphi)
\end{equation}
with $\vvarphi=\big[\phi(-\mu_1/\sigma_1), \phi(-\mu_2/\sigma_2), \ldots, \phi(-\mu_L/\sigma_L)\big]^\top$. By pre-multiplying both sides of~\eqref{eq:vw5} by $\E\{\mPhi_n\}^{-1}$, we arrive at the mean weight error behavior for the $\ell_1$-RLS algorithm:
\begin{equation}
    \label{eq:vw6}
    \begin{split}
    \E\{\hatvw_n\} &= \lambda \E\{\mPhi_n\}^{-1} \E\{\mPhi_{n-1}\} \E\{\hatvw_{n-1}\} \\
    &\quad + \gamma \E\{\mPhi_n\}^{-1} (\1_L - 2\vvarphi)
    \end{split}
\end{equation}
where \eqref{eq:mPhi2} is used to perform the above recursion. Compared to the non-regularized RLS, the second term on the r.h.s. of~\eqref{eq:vw6} corresponds to the bias caused by the sparsity-promoting regularizer. A large $\gamma$ increases the sparsity of the solution while introducing a significant bias. When $\gamma=0$, equation~\eqref{eq:vw6} reduces to the mean weight model of the RLS algorithm.

\subsection{Mean-Square Error Behavior Model}
\label{subsec:MeanSquareAnalysis}

We shall now analyze the $\ell_1$-RLS algorithm in the mean-square error sense. Squaring~\eqref{eq:en1} and taking its expected value, then considering assumption A1 and the statistical property of $z_n$, the mean-square error (MSE) can be expressed as
\begin{equation}
    \label{eq:MSE}
    \E\{e_n^2\} \approx \sigma_z^2 + \tr\{\mR_x \mK_{n-1}\}
\end{equation}
where the second term on the r.h.s. of the above equation denotes the excess mean-square error (EMSE)~\cite{Haykin1991, Sayed2003}. Moreover, the instantaneous mean-square-deviation (MSD) is defined by:
\begin{equation}
    \label{eq:MSD}
    \MSD_n = \E\{\|\hatvw_n\|^2\} = \tr\{\mK_n\}.
\end{equation}
In order to evaluate the EMSE or the MSE, and the MSD, we have to determine a recursion to calculate the correlation matrix $\mK_n$. Post-multiplying~\eqref{eq:vw2} by its transpose, taking the expectation of both sides, and using the statistical property of noise $z_n$, yields
\begin{align}
    \label{eq:mK1}
    \E\{\mPhi_n\hatvw_n & \hatvw_n^\top \mPhi_n \} = \lambda^2 \E\{ \mPhi_{n-1} \hatvw_{n-1} \hatvw_{n-1}^\top \mPhi_{n-1} \} + \sigma_z^2 \mR_x \nonumber \\
    &\quad + \gamma^2 \E\big\{ \sgn\{ \vw_n^\star + \hatvw_{n-1} \} \sgn^\top\{ \vw_n^\star + \hatvw_{n-1} \} \big\} \nonumber \\
    &\quad + \lambda \gamma \E\big\{ \mPhi_{n-1} \hatvw_{n-1} \sgn^\top\{ \vw_n^\star + \hatvw_{n-1} \} \big\} \\
    &\quad + \lambda \gamma \E\big\{ \sgn\{ \vw_n^\star + \hatvw_{n-1} \} \hatvw_{n-1}^\top \mPhi_{n-1} \big\}. \nonumber
\end{align}
Evaluating~\eqref{eq:mK1} is difficult without using some approximations. The following approximations are used in the sequel:
\begin{equation}
    \label{eq:Approx2}
    \E\{ \mPhi_n \hatvw_n \hatvw_n^\top \mPhi_n \} \approx \E\{ \mPhi_n \} \mK_n \E\{ \mPhi_n \},
\end{equation}
\begin{equation}
    \label{eq:Approx3}
    \begin{split}
    &\E\big\{ \mPhi_{n-1} \hatvw_{n-1} \sgn^\top\{ \vw_n^\star + \hatvw_{n-1} \} \big\} \\
    &\qquad\approx \E\{ \mPhi_{n-1} \} \E \big\{ \hatvw_{n-1} \sgn^\top\{ \vw_n^\star + \hatvw_{n-1} \} \big\},
    \end{split}
\end{equation}
\begin{equation}
    \label{eq:Approx4}
    \begin{split}
    &\E\big\{ \sgn\{ \vw_n^\star + \hatvw_{n-1} \} \hatvw_{n-1}^\top \mPhi_{n-1} \big\} \\
    &\qquad\approx \E\big\{ \sgn\{ \vw_n^\star + \hatvw_{n-1} \} \hatvw_{n-1}^\top \big\} \E\{ \mPhi_{n-1} \}.
    \end{split}
\end{equation}
Similar approximations are presented in detail in~\cite{Eweda2020}. The proofs of~\eqref{eq:Approx2}--\eqref{eq:Approx4} are provided in Appendix~\ref{sec:append1}. Simulation results in the sequel will confirm their validity. Using approximations~\eqref{eq:Approx2}--\eqref{eq:Approx4}, expression~\eqref{eq:mK1} can then be written as: 
\begin{align}
    \label{eq:mK2}
    \E\{ \mPhi_n\} &\mK_n \E\{\mPhi_n \} = \lambda^2 \E\{ \mPhi_{n-1} \} \mK_{n-1} \E\{ \mPhi_{n-1} \} + \sigma_z^2 \mR_x \nonumber \\
    & + \gamma^2 \mQ_1 + \lambda \gamma \big[ \E\big\{ \mPhi_{n-1} \} \mQ_2 + \mQ_2^\top \E\{ \mPhi_{n-1} \} \big]
\end{align}
with
\begin{align}
    \label{eq:mQ1}
    \mQ_1 &= \E\big\{ \sgn\{ \vw_n^\star + \hatvw_{n-1} \} \sgn^\top\{ \vw_n^\star + \hatvw_{n-1} \} \big\}, \\
    \label{eq:mQ2}
    \mQ_2 &= \E\big\{ \hatvw_{n-1} \sgn^\top\{ \vw_n^\star + \hatvw_{n-1} \} \big\}.
\end{align}
Pre-multiplying and post-multiplying~\eqref{eq:mK2} by $\E\{\mPhi_n \}^{-1}$ simultaneously, it results that:
\begin{equation}
    \label{eq:mK3}
    \begin{split}
    \mK_n &= \lambda^2 \E\{ \mPhi_n\}^{-1} \E\{ \mPhi_{n-1} \} \mK_{n-1} \E\{ \mPhi_{n-1} \} \E\{ \mPhi_n\}^{-1} \\
    & + \sigma_z^2 \E\{ \mPhi_n\}^{-1} \mR_x \E\{ \mPhi_n\}^{-1} + \gamma^2 \E\{ \mPhi_n\}^{-1} \mQ_1 \E\{ \mPhi_n\}^{-1} \\
    &+ \lambda \gamma \E\{ \mPhi_n\}^{-1} \big[ \E\big\{ \mPhi_{n-1} \} \mQ_2 + \mQ_2^\top \E\{ \mPhi_{n-1} \} \big] \E\{ \mPhi_n\}^{-1}.
    \end{split}
\end{equation}
In order to perform recursion~\eqref{eq:mK3}, we shall now calculate $\mQ_1$ and $\mQ_2$. Before proceeding by using assumption A2, consider two jointly Gaussian random variables $u$ and $v$ defined by:
\begin{equation}
    \label{eq:uv1}
    \left[\!\! \begin{array}{c}
        u \\
        v
    \end{array} \!\! \right] \sim \N \left( \vmu:= \left[\!\! \begin{array}{c}
                                                  \mu_u \\
                                                  \mu_v
                                                \end{array} \!\! \right], \mSigma_{uv}:= \left[ \begin{matrix}
                                                                                              \sigma_u^2 & \rho_{uv} \\
                                                                                              \rho_{uv} & \sigma_v^2
                                                                                            \end{matrix} \right] \right)
\end{equation}
where $\vmu$ and $\mSigma_{uv}$ denote the mean vector and the covariance matrix, respectively. According to Lemma 2 presented in~\cite{Chen2016SPL}, it holds that:
\begin{align}
    \label{eq:Lemma2}
    &\E\big\{\sgn\{u\} \sgn\{v\}\big\} = \Phi\big(\0_2, [\mu_u, \mu_v]^\top, \mSigma_{uv}\big) \nonumber \\
    & + \Phi\big(\0_2, -[\mu_u, \mu_v]^\top, \mSigma_{uv}\big) - \Phi\big(\0_2, [\mu_u, -\mu_v]^\top, \overline \mSigma_{uv}\big) \nonumber \\
    & - \Phi\big(\0_2, [-\mu_u, \mu_v]^\top, \overline \mSigma_{uv}\big)
\end{align}
with
\vspace{-1mm}
\begin{equation}
    \label{eq:mSigma1}
    \overline \mSigma_{uv} = \left[ \begin{matrix}
                                                    \sigma_u^2 & -\rho_{uv} \\
                                                     -\rho_{uv} & \sigma_v^2
                                                   \end{matrix} \right].
\end{equation}
On the one hand, observe that the main diagonal entries $[\mQ_1]_{ii}$ for $1\leq i\leq L$ are equal to $1$. On the other hand, considering assumption A2 and~\eqref{eq:Lemma2}, the off-diagonal entries $[\mQ_1]_{ij}$ for $1\leq i\neq j\leq L$ can be obtained by making the identifications:
\begin{align}
    \label{eq:u}
    [\vw^\star + \hatvw_{n-1}]_i &\to u, \\
    \label{eq:v}
    [\vw^\star + \hatvw_{n-1}]_j &\to v,
\end{align}
with
\vspace{-2mm}
\begin{align}
    \label{eq:mu_u}
    \E\big\{ [\vw^\star + \hatvw_{n-1}]_i \big\} &\to \mu_u, \\
    \label{eq:mu_v}
    \E\big\{ [\vw^\star + \hatvw_{n-1}]_j \big\} &\to \mu_v, \\
    \label{eq:sigma_u}
    \E\big\{ [\hatvw_{n-1}]_i^2 \big\} - \E\big\{ [\hatvw_{n-1}]_i \big\}^2 &\to \sigma_u^2, \\
    \label{eq:sigma_v}
    \E\big\{ [\hatvw_{n-1}]_j^2 \big\} - \E\big\{ [\hatvw_{n-1}]_j \big\}^2 &\to \sigma_v^2, \\
    \label{eq:rho_uv}
    \!\!\!\!\E\big\{ [\hatvw_{n-1}]_i [\hatvw_{n-1}]_j \big\} \!- \E\big\{ [\hatvw_{n-1}]_i \big\} & \E\big\{ [\hatvw_{n-1}]_j \big\}\to \rho_{uv},
\end{align}
where the expectations $\E\big\{ [\hatvw_{n-1}]_i [\hatvw_{n-1}]_j \big\}$ are available from the off-diagonal entries $[\mK_{n-1}]_{ij}$. Furthermore, based on definition~\eqref{eq:uv1} and Lemma 3 proved in~\cite{Chen2016SPL}, it holds that:
\begin{align}
    \label{eq:Lemma3}
    & \E\big\{ u \, \sgn\{v\} \big\} \\
    &= \frac{1}{\sqrt{2\pi a|\mSigma_{uv}|}} \bigg\{ \sqrt{\frac{2\pi}{\theta}} \Big( \mu_u + \frac{c}{a}\mu_v \Big) \Big[ 1 - 2\phi\big( \! - \mu_v \sqrt{\theta} \big) \Big] \nonumber \\
    &- \frac{c}{a} \sqrt{\frac{2\pi}{\theta}} \bigg[ \sqrt{\frac{2}{\pi\theta}} \exp\!\left(\!- \frac{1}{2} \mu_v^2 \theta \right) \! + \mu_v \! \left(1 - 2\phi\big(\! -\mu_v \sqrt{\theta} \big) \right) \! \bigg] \bigg\} \nonumber
\end{align}
where $\theta=b-c^2/a>0$ with
\begin{equation}
    \label{eq:mSigma_inv}
    \mSigma_{uv}^{-1} = \left[ \begin{matrix}
                            a & c \\
                            c & b
                          \end{matrix} \right].
\end{equation}
Likewise, in order to evaluate matrix $\mQ_2$, we can make the following identifications:
\vspace{-2mm}
\begin{align}
     [\hatvw_{n-1}]_i & \to u, \\
     [\vw^\star + \hatvw_{n-1}]_j & \to v,
\end{align}
with
\vspace{-2mm}
\begin{align}
     \E\big\{ [\hatvw_{n-1}]_i \big\} & \to \mu_u, \\
     \E\big\{ [\vw^\star + \hatvw_{n-1}]_j \big\} & \to \mu_v,
\end{align}
for $1\leq i,j \leq L$. Consequently, all the entries of matrix $\mQ_2$ can be determined according to~\eqref{eq:Lemma3} under assumption A2. Using~\eqref{eq:mPhi2}, \eqref{eq:mQ1}, and~\eqref{eq:mQ2}, we can finally perform recursion~\eqref{eq:mK3}, which allows us to characterize the transient mean-square error behavior of the $\ell_1$-RLS algorithm. Note that~\eqref{eq:mK3} reduces to the mean-square weight error model of the RLS for $\gamma=0$.

\section{Numerical Tests}
\label{sec:Tests}

\setcounter{figure}{1}
\begin{figure*}[!htbp]
	\centering
    \subfigure[Empirical vs. theo. evolution for $w_i$ $(\rho=0.6)$.]
	{\includegraphics[trim = 0mm 1.5mm 9mm 6.5mm, clip, width=0.3\textwidth]{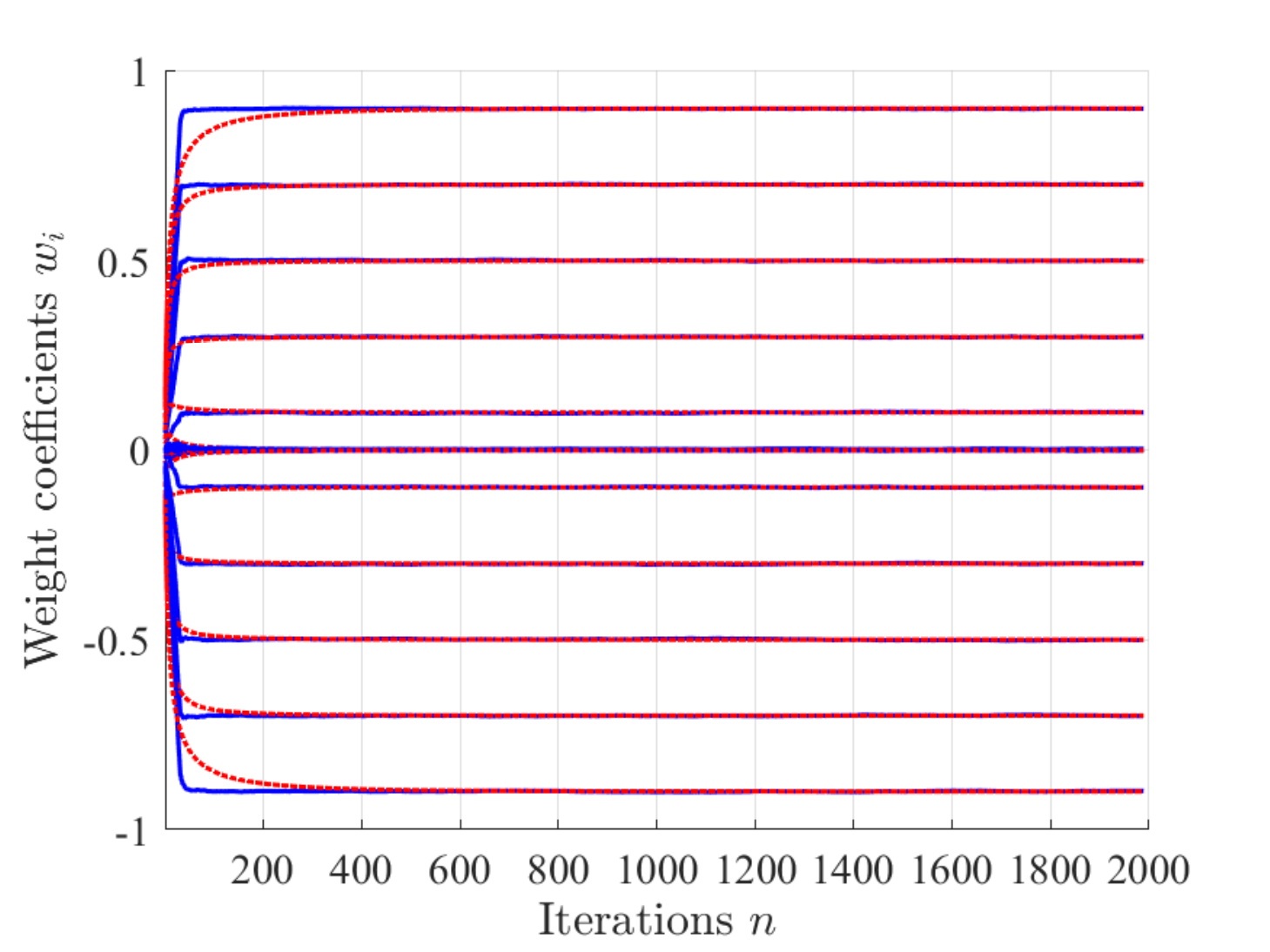}}
    \subfigure[MSE and EMSE curves $(\rho=0.6)$.]
	{\includegraphics[trim = 0mm 1.5mm 9mm 6.5mm, clip, width=0.3\textwidth]{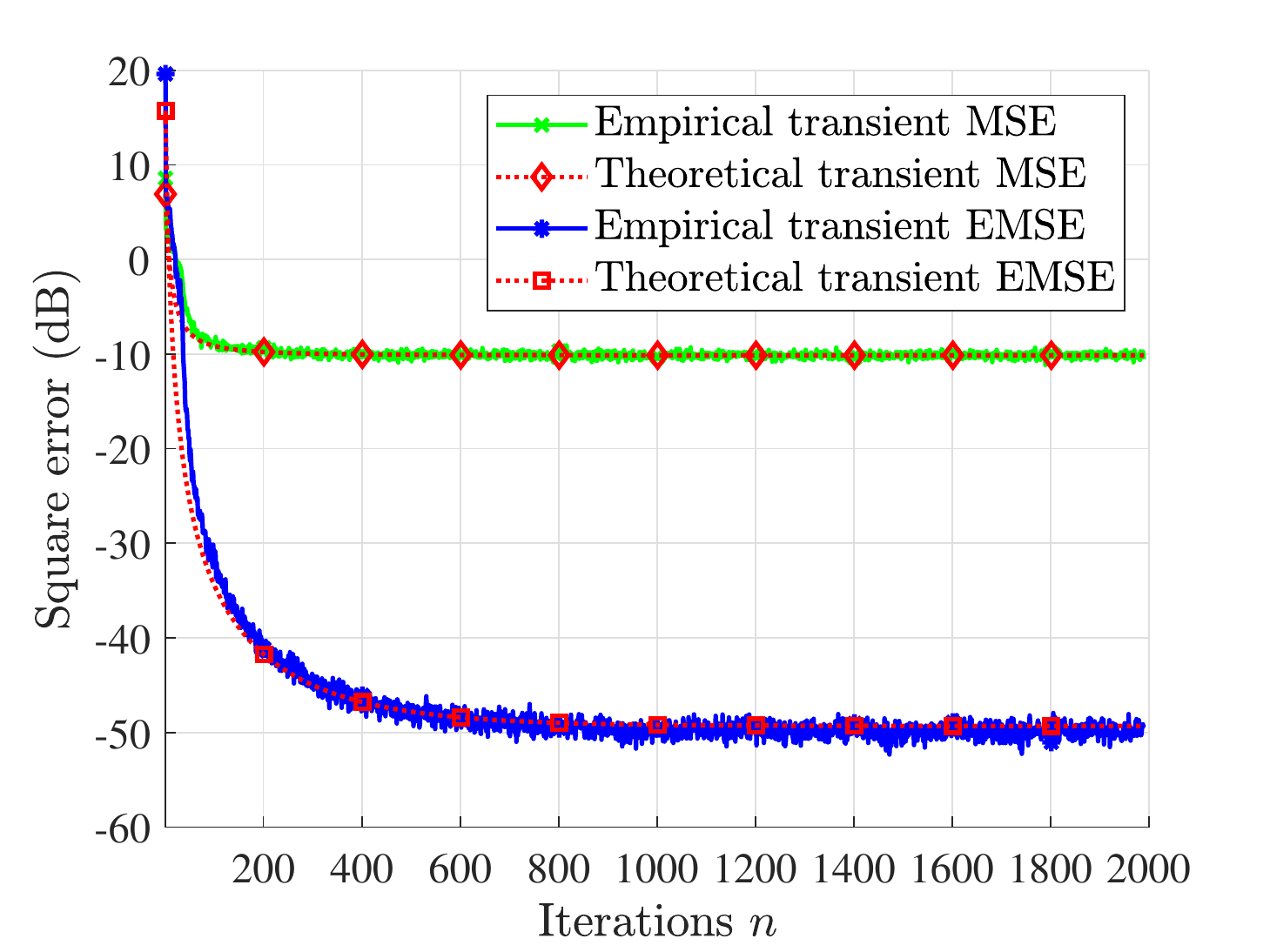}}
    \subfigure[MSD curves.]
	{\includegraphics[trim = 0mm 1.5mm 9mm 6.5mm, clip, width=0.3\textwidth]{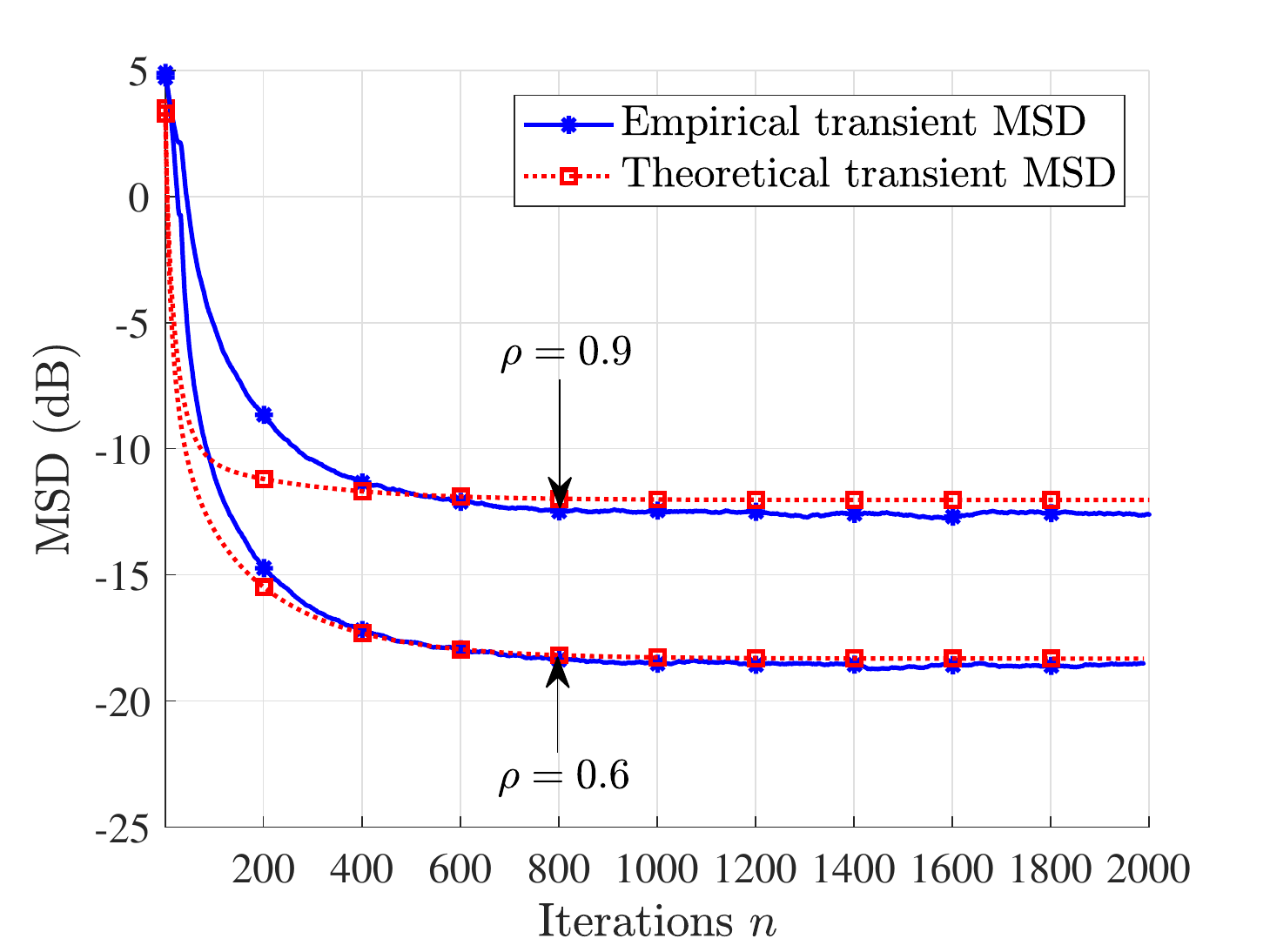}}
    \caption{Comparisons of empirical and theoretical simulation results for the $\ell_1$-RLS algorithm.}
	\label{fig:Results}
\vspace{-4mm}
\end{figure*}

The good performance of the $\ell_1$-RLS algorithm has already been illustrated in the literature by comparing it with related adaptive filters for sparse system identification~\cite{Eksioglu2010, Eksioglu2011IET, Eksioglu2011}. We shall now examine the accuracy of the analytical models derived in this paper, via simulation results. All empirical curves were obtained by averaging over $500$ Monte Carlo runs.

The input signal was generated with a first-order AR model, namely, $x_n = \rho \, x_{n-1} + s_n$, with $\rho$ the correlation factor and $s_n$ a zero-mean white Gaussian random sequence. Its variance was set to $\sigma_s^2=0.64$, in order that the variance of the input signal $x_n$ was $\sigma_x^2=1$. The noise $z_n$ was zero-mean white and Gaussian with variance $\sigma_z^2=0.09$. The optimal weight vector in~\eqref{eq:Model} was set to:
\begin{equation}
    \label{eq:vw_opt}
    \begin{split}
    \vw^\star &= [0.9, \, 0.7, \, 0.5, \, 0.3, \, 0.1, \, \0_{22}^\top, \\
    &\quad\,\, -0.1, \, -0.3, \, -0.5, \, -0.7, \, -0.9]^\top \in \R^{32}.
    \end{split}
\end{equation}
The forgetting factor was set to $\lambda=0.995$, and the regularization parameter was set to $\delta=0.25$. Parameter $\varepsilon$ used to initialize $\mPhi_{-1}=\varepsilon^{-1}\mI_L$ was set to $0.1$. As shown in~\cite{Diniz2013}, note that the bias of $\hatvw_n$ caused by the initialization $\mPhi_{-1}=\varepsilon^{-1}\mI_L$ tends to be negligible as $n\to\infty$ for $\lambda<1$. The weight vector was initialized to zero, that is, $\vw_0=\0_{32}$.
\vspace{-4mm}
\setcounter{figure}{0}
\begin{figure}[!htbp]
	\centering
    \subfigure[$\big{[} {[\hatvw_n]}_2, {[\hatvw_n]}_{10} \big{]}$ at $n=200$.]
	{\includegraphics[trim = 0mm 1.5mm 9mm 6.5mm, clip, width=0.24\textwidth]{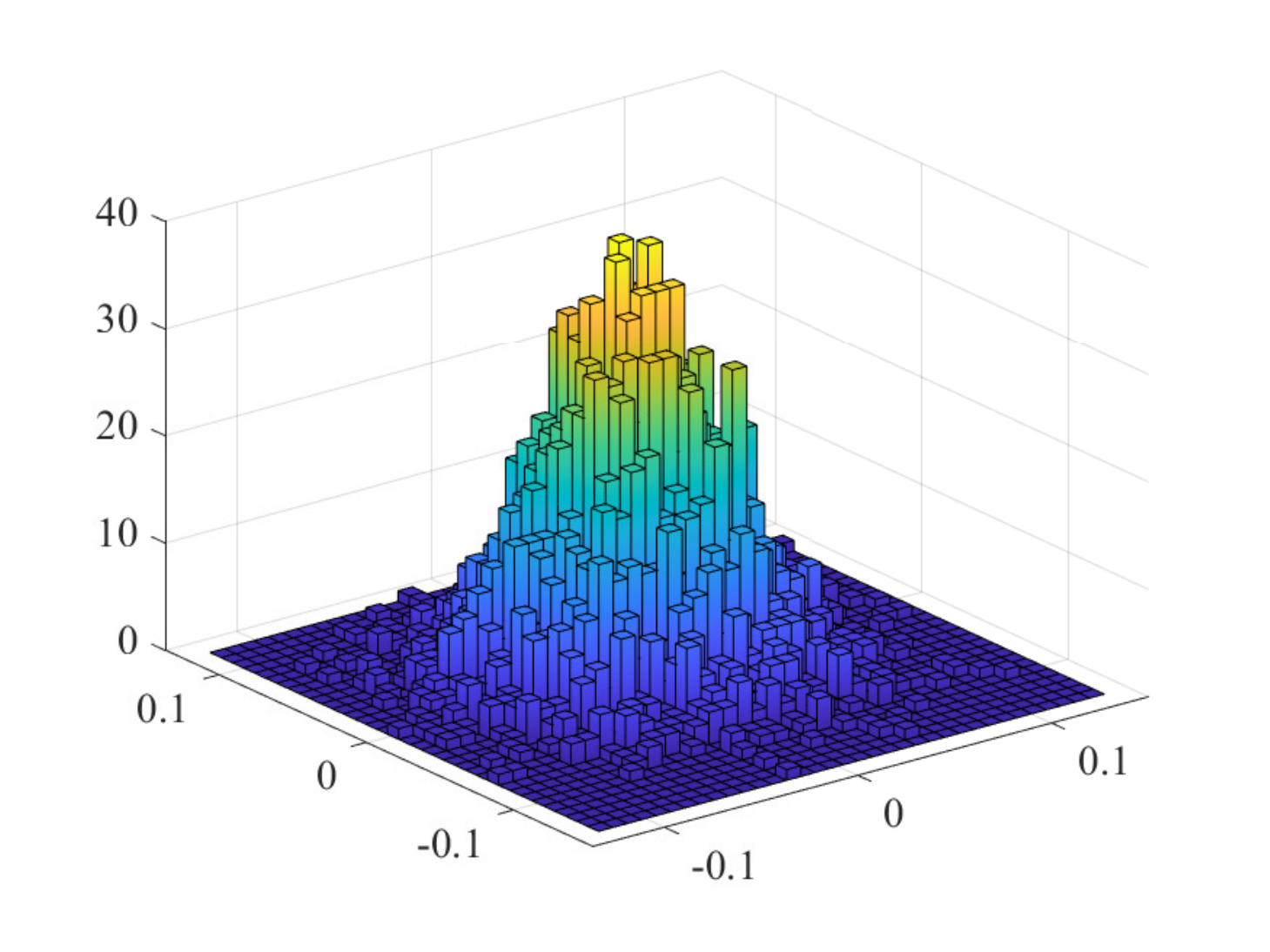}}
    \subfigure[$\big{[} {[\hatvw_n]}_{13}, {[\hatvw_n]}_{25} \big{]}$ at $n=1500$.]
	{\includegraphics[trim = 0mm 1.5mm 9mm 6.5mm, clip, width=0.24\textwidth]{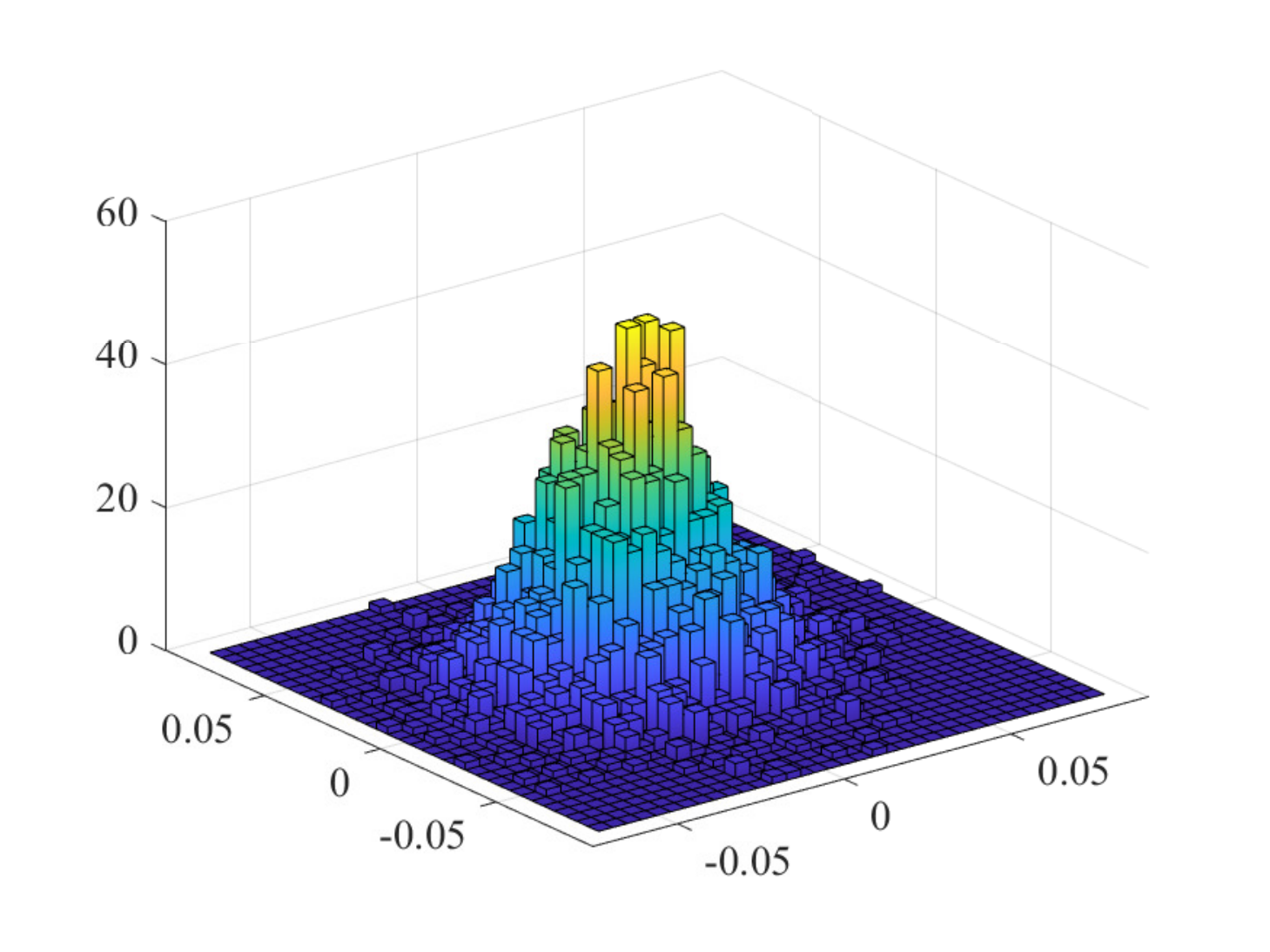}}
    \caption{Histograms of bivariate vector $\big{[} {[\hatvw_n]}_i, {[\hatvw_n]}_j \big{]}$ with $5\times10^3$ samples.}
	\label{fig:Hist3s}
\end{figure}

Two histograms are depicted in Fig.~\ref{fig:Hist3s} for two arbitrarily selected pairs of entries of the weight error vector $\hatvw_n$, i.e., $\big{[} {[\hatvw_n]}_2, {[\hatvw_n]}_{10} \big{]}$ and $\big{[} {[\hatvw_n]}_{13}, {[\hatvw_n]}_{25} \big{]}$ from $5\times10^3$ simulated samples, at time instant $n=200$ and $n=1500$, respectively. These two histograms of bivariate vector $\big{[} {[\hatvw_n]}_i, {[\hatvw_n]}_j \big{]}$ have the required Gaussian-like profiles. To validate that assumption~A2 is feasible, the bivariate normality of these vectors was formally confirmed with Henze-Zirkler's multivariate normality test with significance level $0.05$~\cite{Henze1990, Mecklin2005}. The mean weight behavior is shown in Fig.~\ref{fig:Results}(a). One can observe that all the theoretical curves of weight coefficients $w_i$ predicted by~\eqref{eq:vw6} are generally consistent with those empirical curves including the zero coefficients, which validates the asymptotic unbiasedness of the algorithm. Fig.~\ref{fig:Results}(b) shows that the empirical learning curves of the MSE and EMSE coincide with their theoretical curves obtained from~\eqref{eq:MSE} and~\eqref{eq:mK3}, respectively. Fig.~\ref{fig:Results}(c) shows the good agreement between the empirical learning curve provided by the MSD and its theoretical prediction obtained from~\eqref{eq:MSD} and~\eqref{eq:mK3} for $\rho=0.6$. Note that the mismatch between these two curves, particularly during the initial transient stage, goes larger as $\rho$ increases. Fig.~\ref{fig:Results} illustrates the correctness and accuracy of our analytical models and the necessary approximations~\eqref{eq:Approx2}--\eqref{eq:Approx4} used in the analysis. These models offer an effective means of profoundly understanding the convergence behavior of the $\ell_1$-RLS algorithm in the context of online sparse system identification.

\vspace{-4mm}
\section{Conclusion}
\label{sec:Conclusion}

In this letter, the transient behavior of the $\ell_1$-RLS algorithm was theoretically studied in the mean and mean-square error sense. Simulation results illustrated their accuracy.

\vspace{-2mm}
\appendices
\section{Proofs of Approximations~\eqref{eq:Approx2}--\eqref{eq:Approx4}}
\label{sec:append1}

To prove these three approximations, we denote by $\mDelta_n$ the random fluctuation of matrix $\mPhi_n$ around $\E\{\mPhi_n\}$, namely,
\vspace{-2mm}
\begin{equation}
    \label{eq:mPhi3}
    \mPhi_n = \E\{ \mPhi_n \} + \mDelta_n.
\end{equation}
Then, we have:
\begin{align}
    \label{eq:Approx}
    \E\{ &\mPhi_n \hatvw_n \hatvw_n^\top \mPhi_n \}
    = \E\{ \mPhi_n \} \mK_n \E\{ \mPhi_n \} + \E\{\mDelta_n\hatvw_n \hatvw_n^\top\mDelta_n\} \nonumber \\
    & + \E\{ \mPhi_n\}\E\{\hatvw_n \hatvw_n^\top \mDelta_n\} + \E\{\mDelta_n\hatvw_n \hatvw_n^\top\}\E\{ \mPhi_n\},
\end{align}
\begin{equation}
    \label{eq:Approx5}
    \begin{split}
    &\E\big\{ \mPhi_{n-1} \hatvw_{n-1} \sgn^\top\{ \vw_n^\star + \hatvw_{n-1} \} \big\} \\
    &\qquad= \E\{ \mPhi_{n-1} \} \E \big\{ \hatvw_{n-1} \sgn^\top\{ \vw_n^\star + \hatvw_{n-1} \} \big\} \\
    &\qquad + \E \big\{ \mDelta_{n-1} \hatvw_{n-1} \sgn^\top\{ \vw_n^\star + \hatvw_{n-1} \} \big\},
    \end{split}
\end{equation}
\begin{equation}
    \label{eq:Approx6}
    \begin{split}
    &\E\big\{ \sgn\{ \vw_n^\star + \hatvw_{n-1} \} \hatvw_{n-1}^\top \mPhi_{n-1} \big\} \\
    &\qquad= \E\big\{ \sgn\{ \vw_n^\star + \hatvw_{n-1} \} \hatvw_{n-1}^\top \big\} \E\{ \mPhi_{n-1} \} \\
    &\qquad + \E\big\{ \sgn\{ \vw_n^\star + \hatvw_{n-1} \} \hatvw_{n-1}^\top \mDelta_{n-1} \big\}.
    \end{split}
\end{equation}
We assume that the entries of $\mDelta_n$ are small with respect to those of $\E\{\mPhi_n\}$ based on the fact that~\eqref{eq:mPhi1} shows that $\mPhi_n$ is a low-pass filtering of $\vx_n \vx_n^\top$. The first term on the r.h.s. of each one of the above three equations then dominates the remaining ones, which leads to~\eqref{eq:Approx2}--\eqref{eq:Approx4}.

\newpage


\balance

\bibliographystyle{IEEEtran}
\bibliography{ZA-RLS}

%

%
%
%




\end{document}